\documentclass[11pt,twoside]{article}
\usepackage{asp2010}

\resetcounters

\begin{document}

\title{The acceleration of the Universe in the light of supernovae -- The key role of the Cerro Tololo
Inter-American Observatory}
\author{Mario Hamuy$^1$ 
\affil{$^1$Departamento de Astronom\'ia, Universidad de Chile, Santiago, Chile}}

\begin{abstract}
The discovery of acceleration and dark energy arguably constitutes the most revolutionary discovery
in astrophysics in recent years. Cerro Tololo
Inter-American Observatory (CTIO) played a key role in this amazing discovery through three
systematic supernova surveys organized by staff astronomers: the ``Tololo Supernova Program`` (1986-2000), 
the Cal\'an/Tololo Project (1989-1993), and the ``High-Z Supernova Search Team'' (1994-1998). CTIO's
state of the art instruments also were fundamental in the independent discovery of acceleration by the
``Supernova Cosmology Project'' (1992-1999). Here I summarize the work on supernovae carried out from CTIO that led to
the discovery of acceleration and dark energy and provide a brief historical summary on the use of 
Type Ia supernovae in cosmology in order to provide context for the CTIO contribution. 
\end{abstract}

\section{Introduction}

We live in an accelerating Universe due to a misterious dark energy that 
comprises ~70\% of the Universe. While this paradigm is quite familiar to us
today, it was virtually unimaginable until a few years. How did we get to this
intriguing situation? A fundamental ingredient in this story was the 
discovery and precise measurement of Type Ia supernovae (SNe) done at 
CTIO over the last 27 years. In this paper
I summarize the work on SNe carried out from CTIO that led to 
the discovery of acceleration and dark energy (section \ref{section2})
and provide a brief historical summary on the use of SNe Ia in cosmology in order to
provide context for the CTIO contribution (section \ref{section3}).
Finally I present a summary of the key role played by the observatory in this amazing
journey, which represents a truly collaborative effort by US and Chilean
astronomers that goes back to the very same establishment of CTIO in Chile in the early 1960's.

\section{Supernova Science at CTIO Over Time}
\label{section2}

\subsection{The Tololo Supernova Program}

The introduction of CCDs to CTIO nearly coincides with the first systematic investigations on SNe at the
observatory. M. Phillips and N. Suntzeff were the first staff astronomers that initiated
such studies with the bright and nearby SN~1986G in the galaxy Centaurus A, the first SN ever measured with a CCD.
Unprecedented well-sampled $BV$ lightcurves and optical spectroscopy revealed high obscuration by dust in SN 1986G
and, more generally, the existence of intrinsic differences in the properties of SNe Ia \citep{phillips87}.
A strong boost to SN science worldwide came one year later from Las Campanas Observatory with the serendipitous
discovery by I. Shelton and O. Duhalde of SN~1987A, the first naked-eye
SN since Kepler's SN in 1604. An intensive photometric and spectroscopic follow-up campaign was organized by
CTIO staff astronomers, including M. Hamuy who had just joined CTIO (only two days after Shelton's discovery). 
A first summary of this effort was presented in the CTIO 25th anniversary symposium  by \citet{elias88}, which taught
us that: (1) contrary to expectations, SN 1987A's progenitor was a compact blue supergiant star;
(2) the late time light curve was powered by the radioactive decay of $^{56}$Co into $^{56}$Fe;
(3) dust formation took place in the SN ejecta $\sim$500 days after explosion, among many other
unveiling features. The study of SNe from CTIO gradually took the form of a regular program focused on obtaining detailed
digital observations of bright and nearby SNe of all types. Although this project never received a formal
name, hereafter I will refer to it as the ``Tololo Supernova Program'', led by N. Suntzeff and M. Phillips. Many objects were observed as part
of this project in the following years, among which are some emblematic Type Ia SNe: SN 1989B, SN 1990N, SN 1991T, SN 1991bg, 
and SN 1992A. One of the main results of this effort is the establishment of a 
correlation between decline rate and peak luminosity for SNe Ia, a.k.a. ``Phillips relationship''.
Fig. 1 shows a remake of the original figure based on 9 SNe: 6 of the emblematic objects from the ``Tololo Supernova Program'',
SN 1980N and SN 1981B in Fornax A from the El Roble survey \citep{hamuy91}, and one object (SN 1971I) taken from the literature \citep{deming73}.
This empirical relationship constitutes the basic underlying idea today behind precision cosmology from SNe Ia.

\begin{figure} [h]
\begin{center}
\includegraphics[height=7in,width=7in,angle=0,scale=0.5]{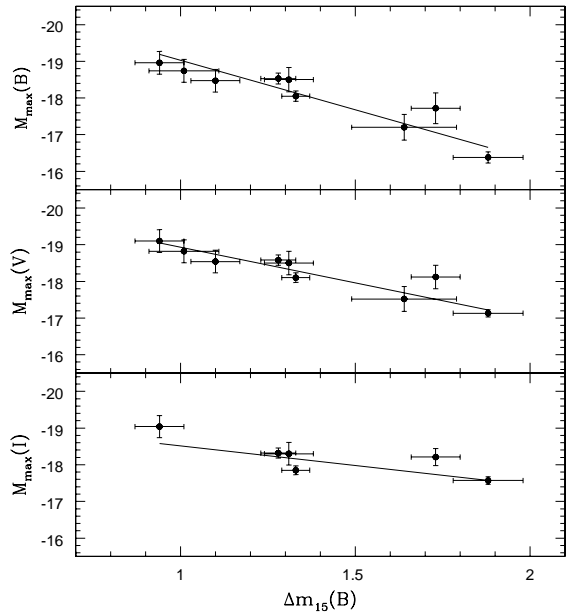}
\caption{Decline rate-peak luminosity relationship for 9 nearby SNe Ia analyzed by \citet{phillips93}.}
\end{center}
\label{fig1}
\end{figure}

\subsection{The Cal\'an/Tololo low-z Project (1989-1993)}

Soon after the tenth Santa Cruz summer workshop on SNe, timely organized by S. Woosley in 1989 to
discuss the first results on SN~1987A, a new project was born at CTIO: a systematic survey of ``faint'' 
Type Ia SNe in the Hubble flow in order to study the usefulness of these objects as distance indicators. 
The project, initially inspired by the work presented by B. Leibundgut\footnote{at the time a PhD student at the University of Basel 
under the guidance of G. Tammann.} on SNe Ia \citep{leibundgut91} in the just-mentioned
workshop, was a collaboration between astronomers from 
CTIO (M. Hamuy, M. Phillips, N. Suntzeff, R. Schommer, C. Smith, L. Wells) and Cerro Cal\'an of the University of Chile
in Santiago (R. Antezana, L. Gonz\'alez, P. Lira, J. Maza, M. Wischnjewsky). 
It consisted in a photographic search for southern SNe using the CTIO Schmidt Camera -- a reincarnation of
the Cerro El Roble SN photographic search carried out by J. Maza and collaborators between 1979-1984 with
the University of Chile's Makstuvov telescope \citep{maza81} --
and a followup program with the 0.9m, 1.5m and 4m CTIO telescopes aimed at getting modern CCD optical photometry 
and spectroscopy of such objects \citep{hamuy93}. The first tests with the Schmidt telescopes began in 1989.
The effort paid off and in the following four years the Cal\'an/Tololo project had discovered 50 SNe 
(25\% of all discoveries worldwide), 32 of which proved to be SNe Ia.
The main results of the Cal\'an/Tololo project can be summarized as follows:

\begin{itemize}

\item The discovery of 32 SNe Ia SNe in the Hubble flow, a world
record at the time, and succeeding where other astronomers had previously failed.

\item The recording of the most precise light curves ever obtained at that
time for 29 SNe Ia, thanks to the recently adopted revolutionary CCD technology in
astronomy (Fig. \ref{fig2}; left).

\begin{figure} [h]
\plottwo{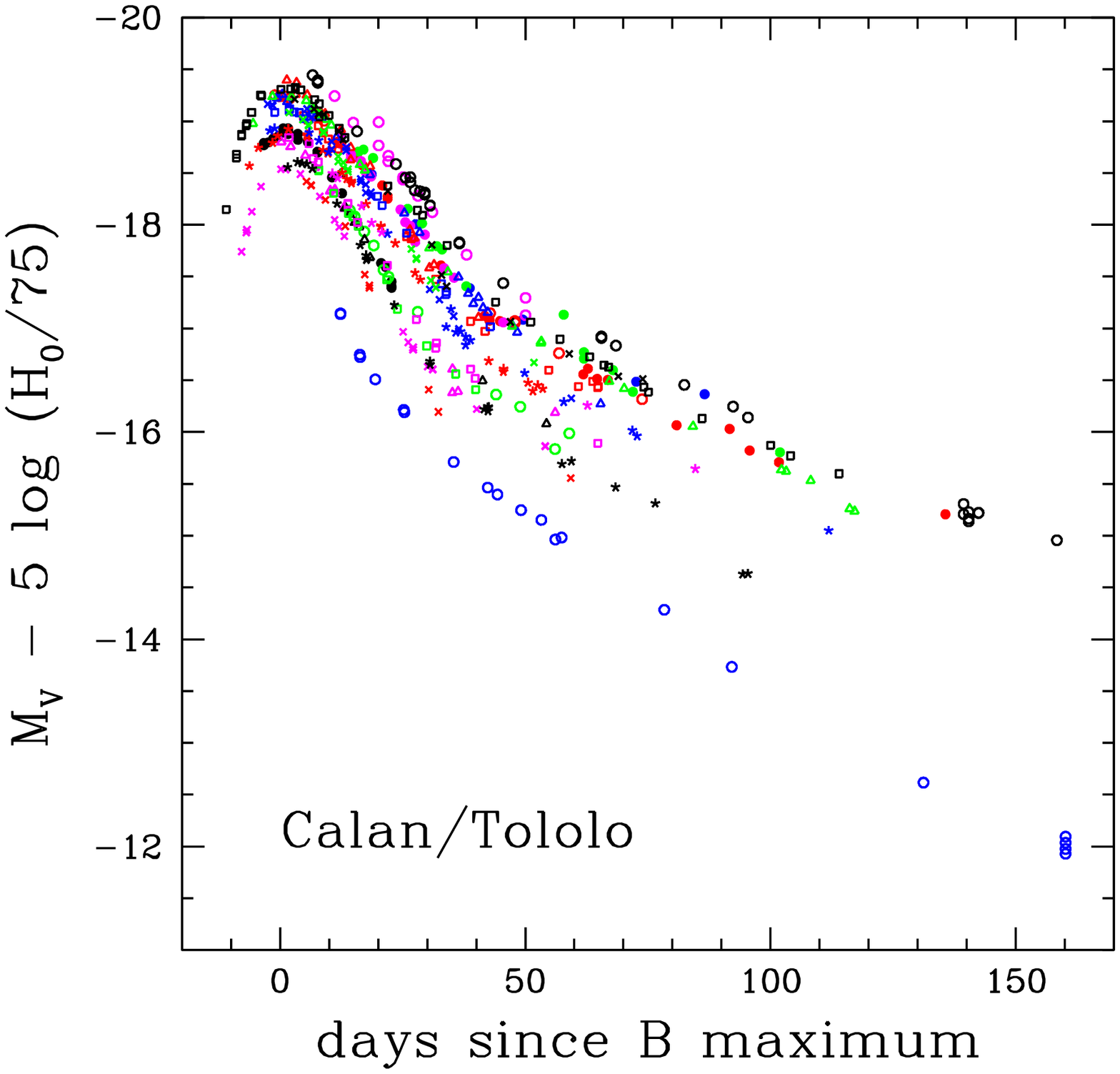}{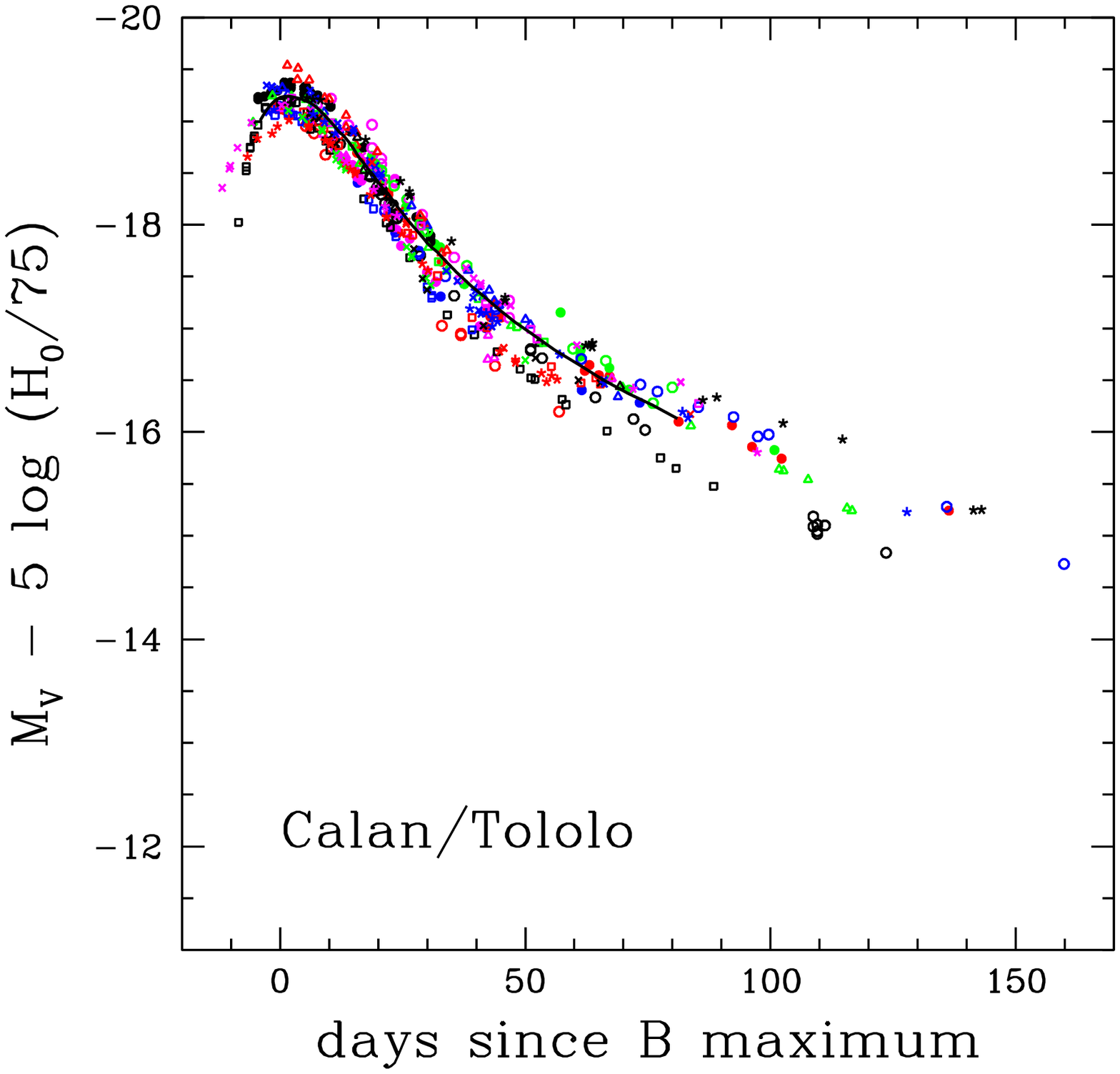}
\caption{(left) $V$-band lightcurves for 29 Cal\'an/Tololo SNe Ia. (right) $V$-band corrected lightcurves for the same sample.}
\label{fig2}
\end{figure}

\item The proof that SNe Ia were not perfect standard candles, 
which can be clearly seen in the photometric diversity shown in Fig. \ref{fig2} (left),
both in terms of peak brightness and light curve width.

\item The demonstration that Phillips (1993) relation was qualitatively
correct. As Fig. \ref{fig3} (left) shows it, the Cal\'an/Tololo SN sample confirmed
the decline rate-peak luminosity relationship for nearby SNe, but with
a shallower slope.

\begin{figure} [h]
\plottwo{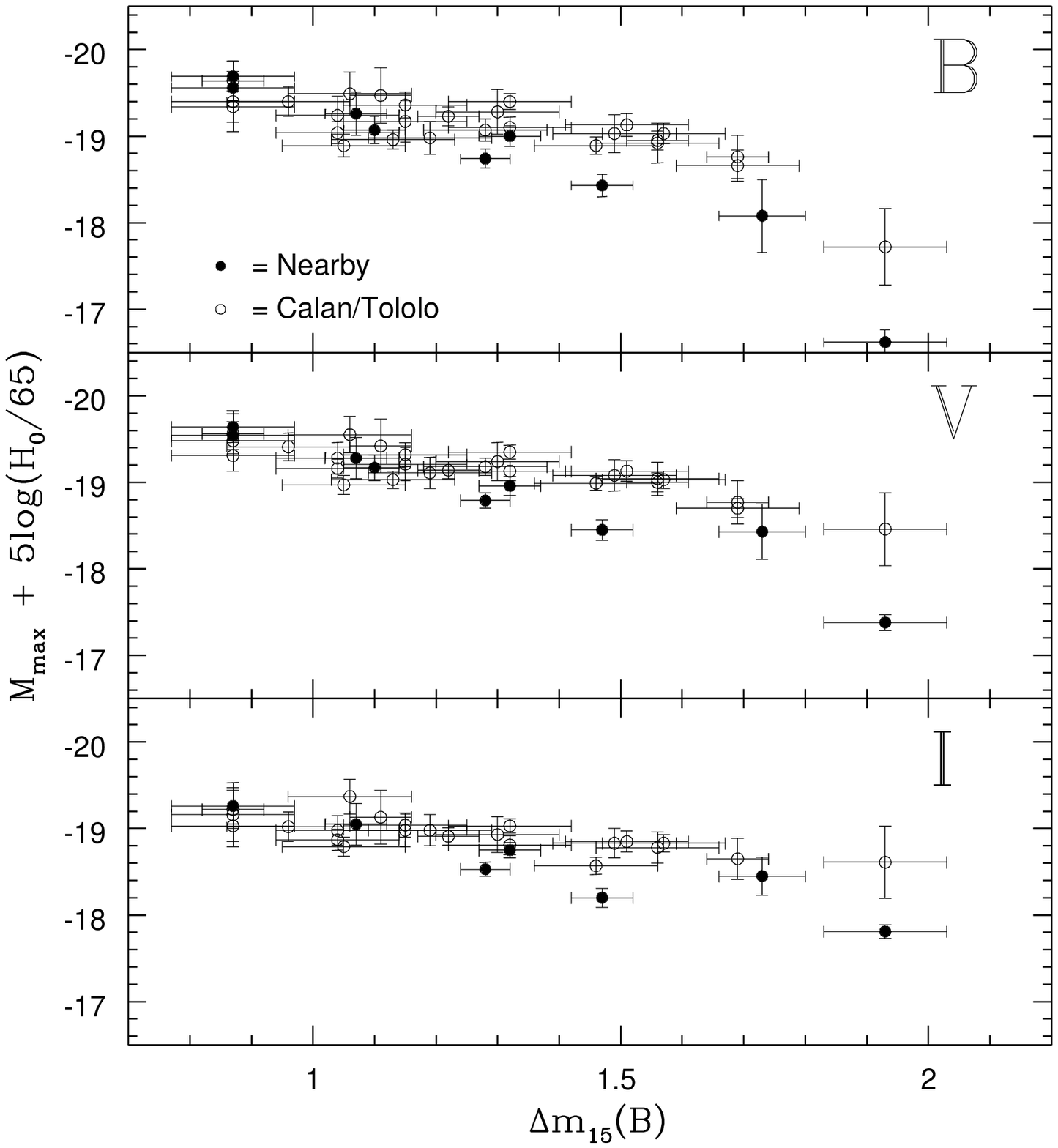}{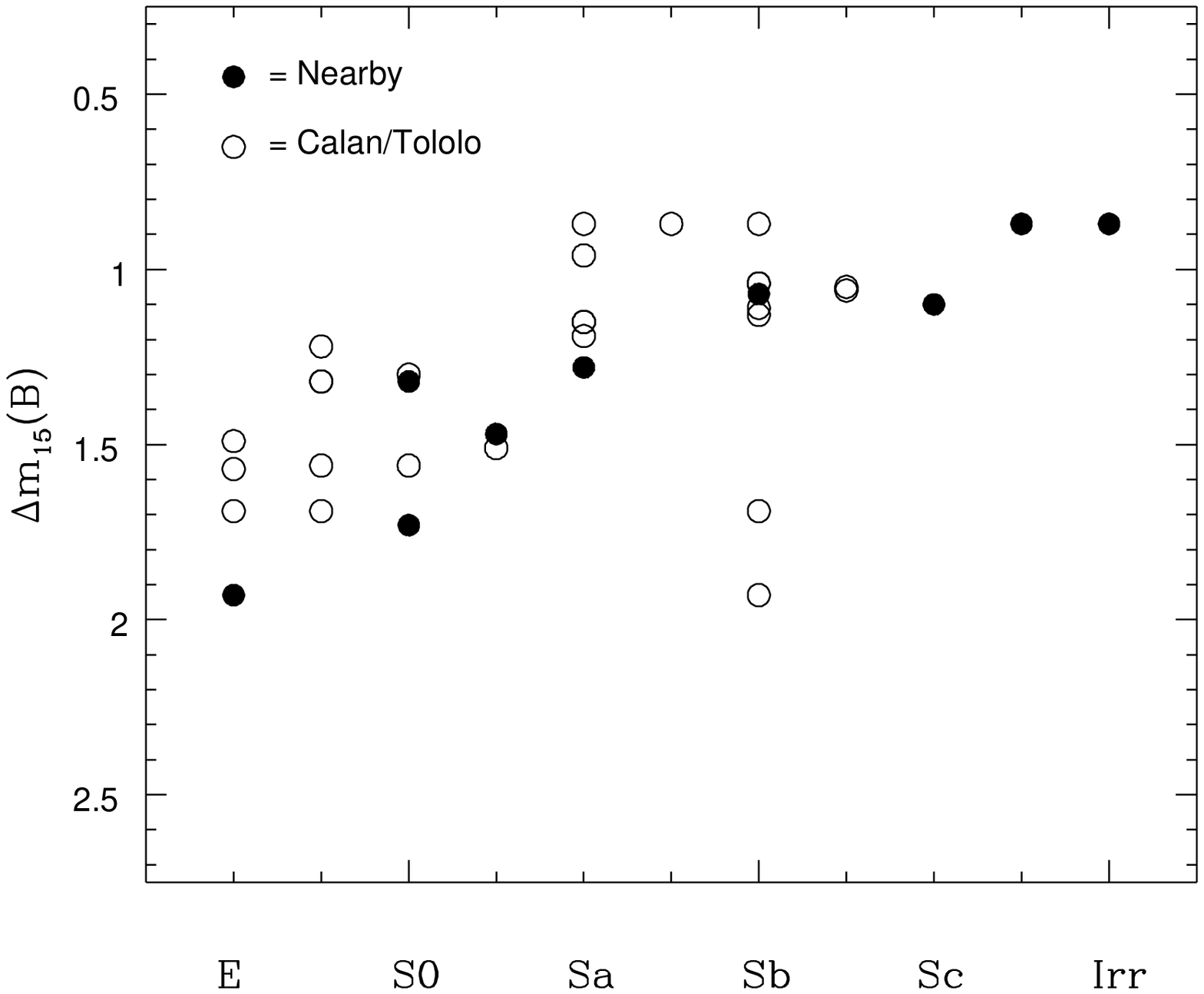}
\caption{(left) Peak magnitude -- Decline rate relationship for the nearby and Cal\'an/Tololo SNe.
(right) Decline rate dependence with host galaxy type \citep[from][]{hamuy96a}.}
\label{fig3}
\end{figure}

\item The dependence of SN Ia luminosities with galaxy types. As shown
in Fig. \ref{fig3} (right), early type galaxies preferentially host fast declining SNe Ia 
(with higher $\Delta m_{15}$ values), i.e. intrinsically fainter SNe.

\item The establishment of a method to correct the SN luminosities for
host-galaxy reddening (a.k.a. ``Lira law''). The method makes use of the fact that SNe Ia
evolve to a common B-V color 30 days after maximum, despite showing remarkable
differences at earlier times (see Fig. \ref{fig4}; left).

\begin{figure} [h]
\plottwo{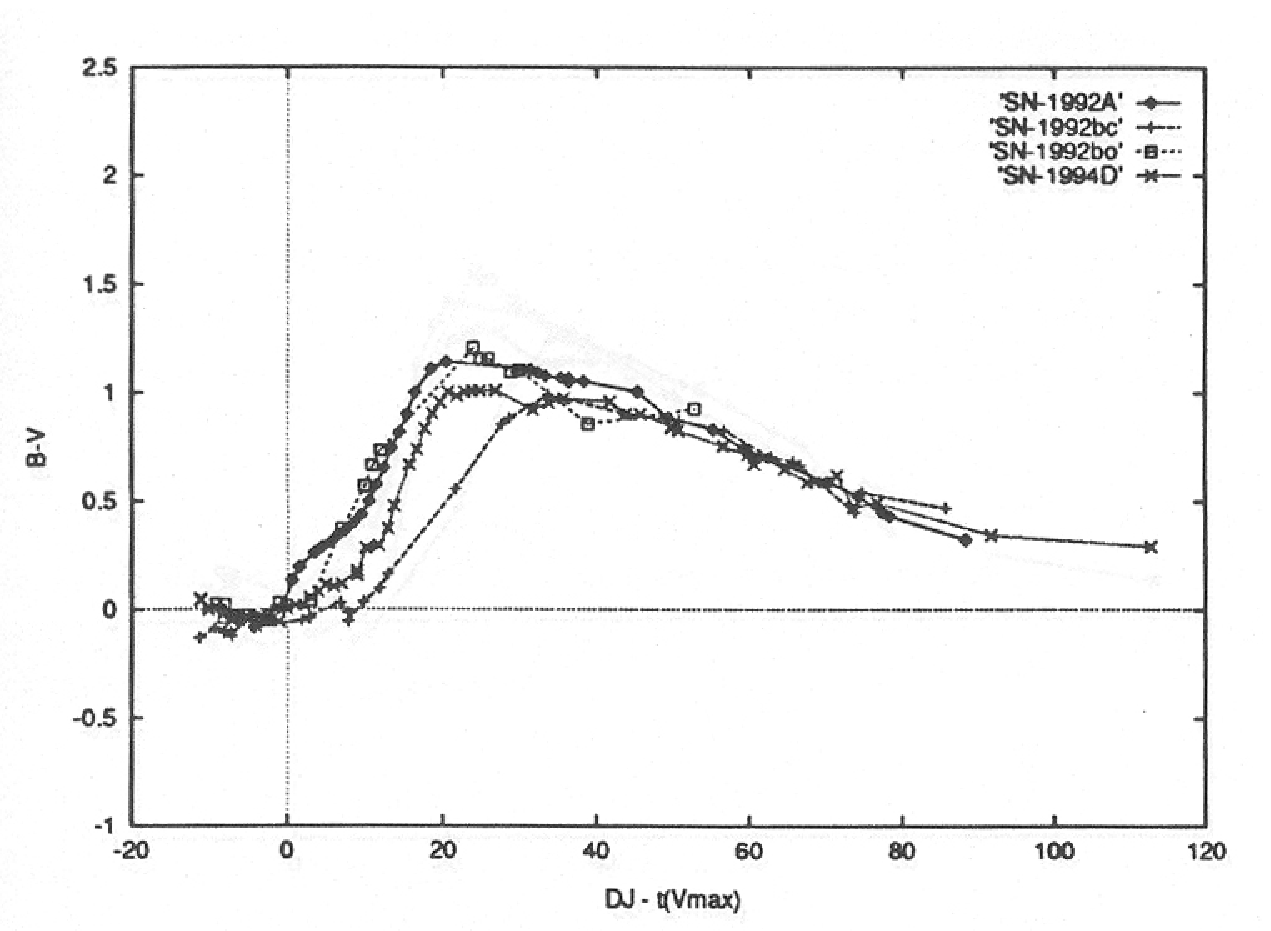}{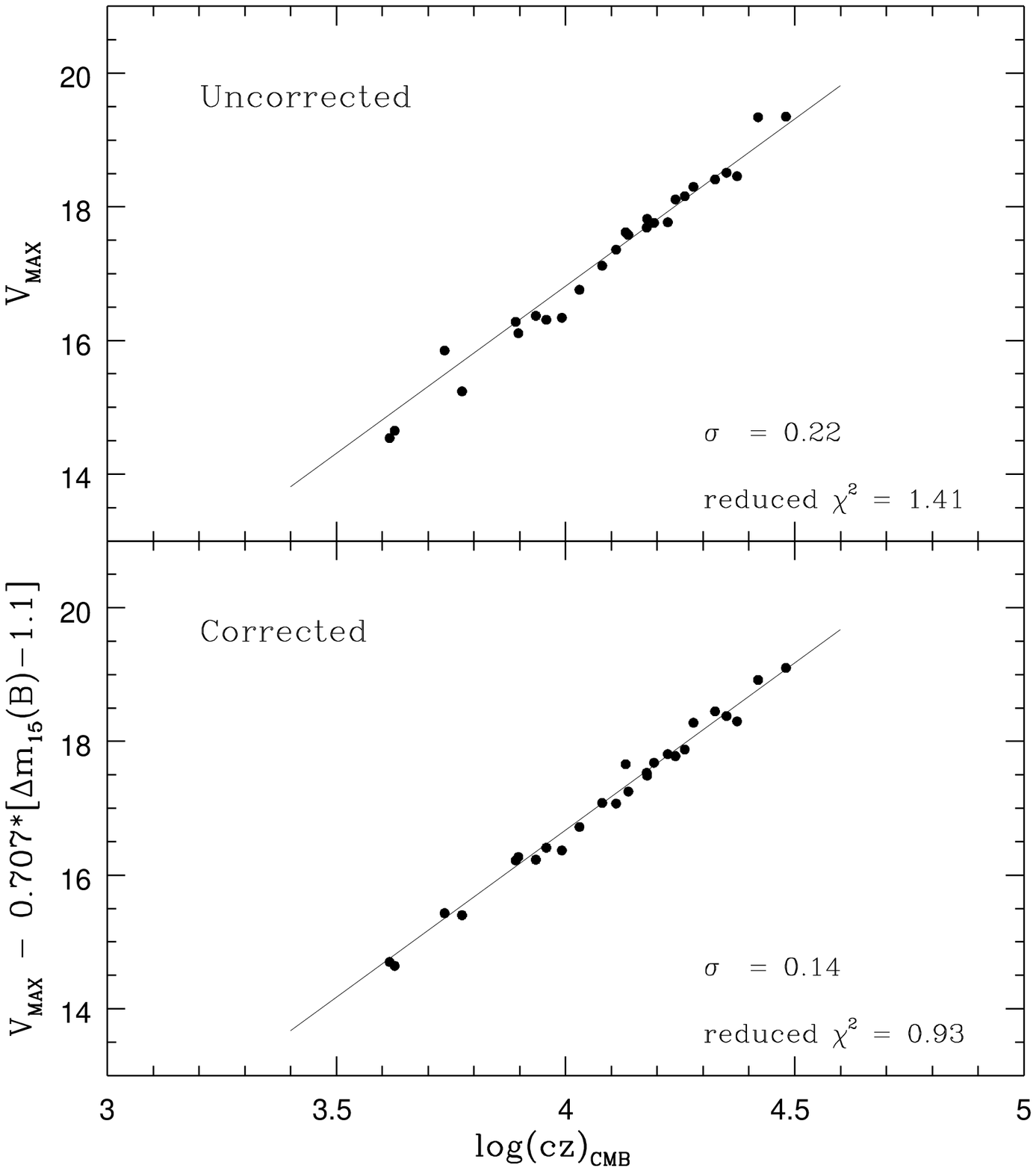}
\caption{(left) B-V color curves for 4 nearby SNe Ia, showing the ``Lira Law'' at late times \citep[from][]{lira95}. 
(right) $V$-band Hubble diagram for the Cal\'an/Tololo SNe Ia.
The upper panel shows magnitudes uncorrected for decline rate, while the lower panel
is based on corrected magnitudes \citep[from][]{hamuy96b}.}
\label{fig4}
\end{figure}

\item The most precise calibration of the SN luminosities at the time. As Fig. \ref{fig2} (right)
shows it, a very uniform photometric standard lightcurve is obtained after applying the  
Cal\'an/Tololo decline rate-peak luminosity relationship. Note that the corrected $V$-band peak
magnitudes have not been forced to coincide, i.e, the resulting scatter reflects the true dispersion
in the corrected luminosities. 

\item The establishment of key tools to measure distances with a precision $\sim$7\%,
never reached before, as can be seen in the $V$-band Hubble diagram shown
in Fig. \ref{fig4} (right).

\end{itemize}

\subsection{The high-z Projects -- The Discovery of Acceleration}

The successful results the Cal\'an/Tololo project began to unfold by 1993, at
which time (September 1993) the team submitted a telescope proposal to begin a new 
SN search with the Schmidt telescope, but this time using a new instrument: a
digital camera equipped with a TEK 2048 CCD. The goal was to ``investigate
the next stage in the Cal\'an/Tololo survey: a deeper CCD-based survey to z$\sim$0.25
which will eventually prove useful for estimating $q_0$''. The observations were
carried out in early 1994 but the analysis (image subtraction) proved difficult
due to the coarse pixel scale of the CCD system and an innaccurate telescope
pointing system. 

The idea of extending Cal\'an/Tololo to higher
redshifts was followed up later that year by N. Suntzeff and B. Schmidt, a
recent PhD graduate of Harvard University. A first meeting took place in the 
CTIO headquarters in La Serena, which led to the establishment of an expanded 
collaboration for the search of high-z SNe with the CTIO 4m telescope.
The original members of the ``High-Z Supernova Search Team'' (HZT) included Cal\'an/Tololo researchers
N. Suntzeff (held responsible for organizing the project), C. Smith, R. Schommer, 
M. Phillips, M. Hamuy, R. Avil\'es, and J. Maza, B. Schmidt (who became a few months
later director of the project by a vote of the HZT members, while holding a postdoctoral
position at Mount Stromlo), A. Riess and R. Kirshner from Harvard,
and J. Spyromilio and B. Leibundgut from the European Southern Observatory.
A first telescope proposal titled ``Pilot Project
to Search for Distant Type Ia Supernovae''\footnote{$http://en.wikipedia.org/wiki/File:Pages\_from\_Prop1994.jpg$}
was submitted in 1994 September 29, 
requesting 4 nights during March-April 1995 with the Tek2048 PFCCD with the purpose
to search for SNe Ia at z$\sim$0.3-0.5, using the same observational technique of the
Cal\'an/Tololo survey (imaging fields before and after a dark run) to ensure the
discovery of young SNe. The efforts promptly paid off with the discovery of SN~1995K 
at z=0.48 \citep{phillips95,schmidt98} thanks to an innovative automated search software developed
by the team\footnote{Based on a revolutionary image subtraction code previously developed at CTIO by A. Phillips.}
and to the ESO collaborators for getting a SN spectrum clean of host galaxy light.
The discovery of high-z SNe became much more efficient with the 
deployment in 1997 of the BTC (Big Throughput Camera; a mosaic of
four SITe 2048 CCDs) on the 4m telescope, and later on with the
upgrade from the PFCCD to the NOAO MOSAIC Camera.
The successful survey promptly led to the use
of the HST and other ground based telescopes such as the CFHT, Keck, NTT, among many others. 

\begin{figure} [h]
\plottwo{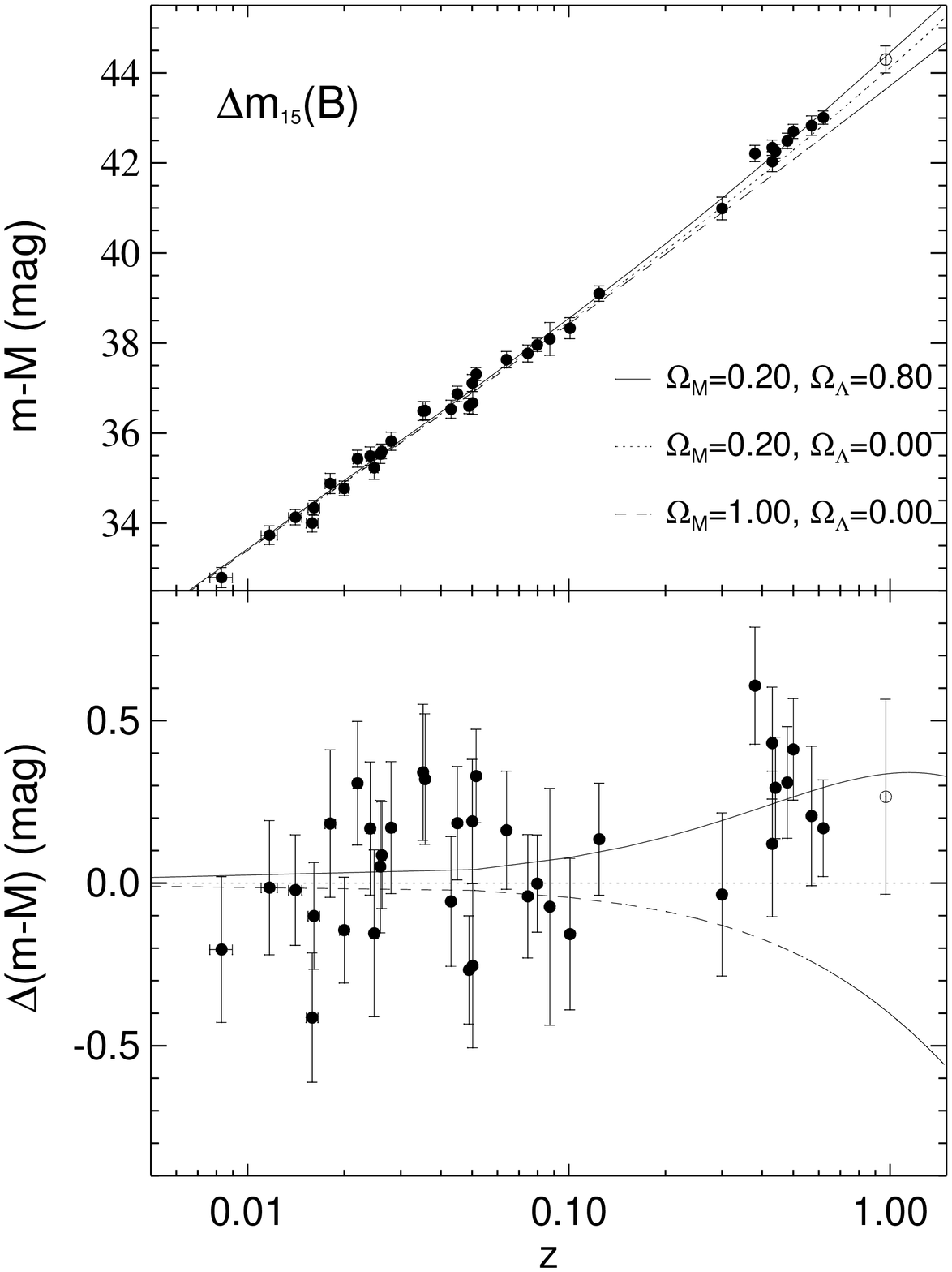}{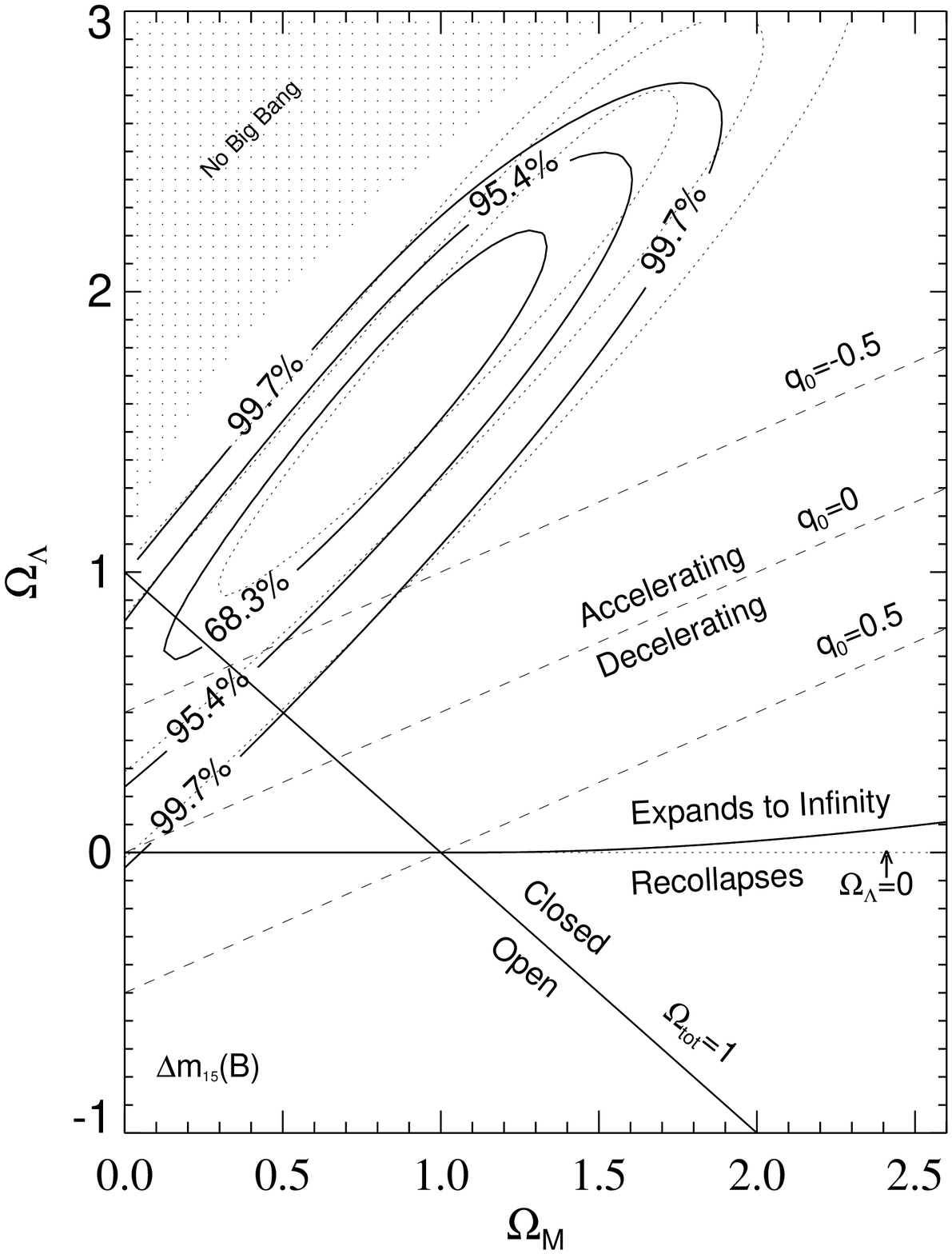}
\vspace{0.5cm}
\caption{(left) SNe Ia Hubble diagram from the HZT. (right) Joint confidence intervals for ($\Omega_M$,$\Omega_{\Lambda}$) from the HZT \citep[both from][]{riess98a}.} 
\label{fig5}
\end{figure}

A first high-z Hubble diagram was published by the HZT in 1998 based on four SNe with z=0.5-1.0 \citep{garnavich98},
leading to the conclusion that ``Either the universe is open or flat; if flat, then the
cosmological constant makes a considerable contribution'', suggesting that $q_0<0$, i.e.
that the universe could be accelerating. The confirmation of this bold conclusion would come the same
year but this time based on an expanded set of 14 high-z and 34 nearby SNe Ia \citep{riess98a}. 
The resulting Hubble diagram (Figure \ref{fig5}, left) revealed that the high-z SNe appeared $\sim$10-15\% 
more distant than expected for a low-mass density universe, implying that the universe is undergoing
an accelerated expansion owing to a positive cosmological constant, i.e. $q_0<0$ and $\Omega_{\Lambda}>0$ 
at the 3$\sigma$ confidence level. The joint confidence intervals for ($\Omega_M$,$\Omega_{\Lambda}$) 
are reproduced in Figure \ref{fig5} (right), which clearly show that the data are inconsistent with
a $\Omega_{\Lambda}=0$ universe. 

\begin{figure} [h]
\center
\includegraphics[height=7in,width=9in,angle=0,scale=0.5]{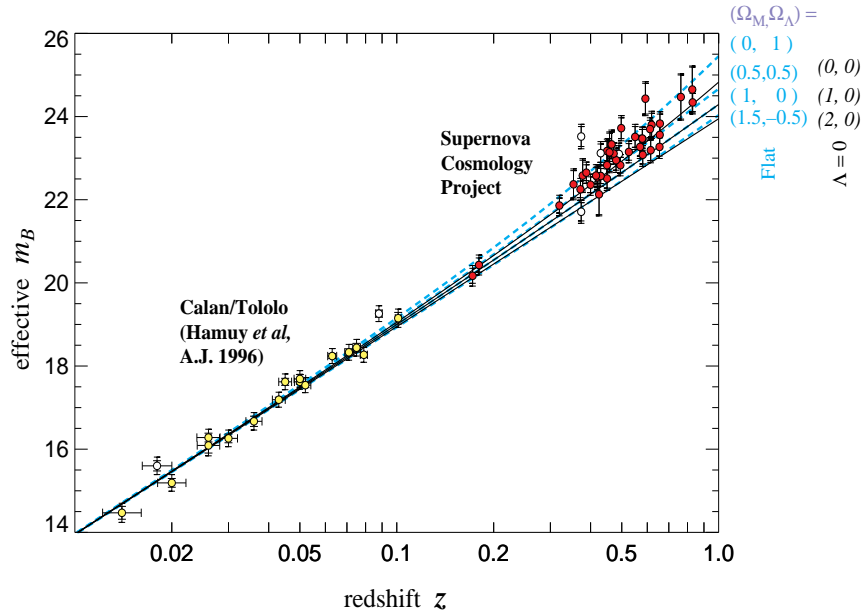}
\caption{SNe Ia Hubble diagram from the SCP \citep[from][]{perlmutter99}.} 
\label{fig6}
\end{figure}

\begin{figure} [h]
\center
\includegraphics[height=6in,width=6in,angle=0,scale=0.5]{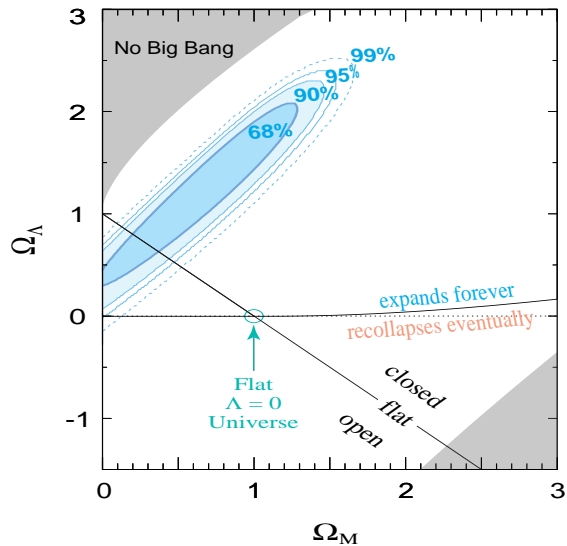}
\caption{Joint confidence intervals for ($\Omega_M$,$\Omega_{\Lambda}$) from the SCP \citep[from][]{perlmutter99}.} 
\label{fig7}
\end{figure}

This surprising result was initially taken with caution by the scientific community. However, the 
independent measurement reported in 1998 by a competing team headed by Saul Perlmutter from
Lawrence Berkeley Laboratory (LBL) lent significant credibility to this finding. The 
``Supernova Cosmology Project'' (SCP) had begun in 1988 with the purpose to measure $q_0$ using SNe Ia.
After four years of hard work, the first high-z SN was found in 1992 and by 1998 more that 75 SNe Ia at z=0.18-0.86
had been discovered and studied by the CSP. The results reported simultaneously with the \citet{riess98a} paper but
published in 1999 were based on 
42 high-z SNe Ia, all discovered with the PFCCD, MOSAIC, and the BTC on the CTIO 4m telescope. The photometric
followup was carried out using the CTIO 4m, WIYN 3.6m, ESO 3.6m, INT 2.5m, WHT 4.2m telescopes,
while the classification spectra came from the Keck 10m and ESO 3.6m telescopes. The Hubble
diagram for these 42 high-z SNe is shown jointly with a subset of 18 Cal\'an/Tololo low-z SNe
in Fig. \ref{fig6}. Again, the high-z SNe appeared too far away, making it necessary
to fit the data with a cosmological constant. The confidence regions for ($\Omega_M$,$\Omega_{\Lambda}$)
from SCP, reproduced in Fig \ref{fig7}, demonstrated that the data were inconsistent
with $\Omega_{\Lambda}$=0 with a confidence of 99\%. Assuming a flat universe ($\Omega_M$+$\Omega_{\Lambda}$=1), 
the cosmological constant turned out to be the dominant energy component of energy with a density
$\Omega_{\Lambda}\sim$70\%.

The discovery of the accelerating universe was initially attributed to a comological constant.
However, since other types of energy components, globally known as ``dark energy'', could also
cause acceleration, a new SN search project was launched in 2002 with the new 
MOSAIC-II camera (a mosaic of 8 2048 SITe CCDs) on the CTIO 4m Blanco telescope, 
with the specific purpose to set constraints on the dark energy equation-of-state parameter, 
$w=P/(\rho c^2)$. The ESSENCE project led by C. Stubbs was a reincarnation of 
the HZT expanded to new members. Using 60 high-z SNe discovered between 2002-2005, the ESSENCE
team reported a value of $w$=-1.05 $\pm$ 0.13, i.e. the dark energy responsible for
the acceleration was fully consistent with a cosmological constant ($w$=-1) \citep{woodvasey07}.

The discovery of the accelerating universe in 1998 meant not only a revolution 
in astronomy (unveiling 70\% of the universe that had remained hitherto unnoticed), but also
presented a huge challenge to physics in order to explain the nature of the "dark energy". 
Not surprisingly, the Royal Swedish Academy of Sciences decided to award the Nobel 
Prize in Physics 2011 to S. Perlmutter, B. Schmidt, and A. Riess "for the discovery
of the accelerating expansion of the Universe through observations of distant supernovae".

\section{A brief history on the role of Type Ia Supernovae in Cosmology}
\label{section3}

The discovery of acceleration through SNe Ia is the pinnacle of a series of
scientific efforts, that goes back to the pioneering work of Baade \& Zwicky who recognized in 1938 
that SNe constituted a new class of very luminous astrophysical objects \citep{baade38,zwicky38}.
Analyzing a sample of 18 SNe \citet{baade38} determined a mean absolute photographic
magnitude of $M_{max}= -14.3 \pm 0.42$ ($\sigma$=1.1 mag), thus
revealing the high luminosities of the SN class \citep[note that the basic Type I and II
subclasses were not recognized until the work by][]{minkowski41}. Although
the dispersion was high, this study already showed the potential of SNe as distance
indicators. 

A recalibration of the luminosities of SNe was performed
by \citet{vandenbergh60} who found a mean maximum-light magnitude 
$M_{pg} = -18.7 \pm 0.3$ ($\sigma$=1.1 mag) for SNe I and $M_{pg} = -16.3 \pm 0.3$ ($\sigma$=0.7 mag) for SNe II.
The large difference in absolute magnitude between \citet{baade38} and \citet{vandenbergh60}
was due to calibration differences and the fact that no distinction among
different SN types had been recognized at the time of Baade's work. The dispersion was still 
high for precise distance measurements but, as noted by \cite{vandenbergh60}, the photographic magnitudes 
had large uncertainties, due to poor photometric calibrations, potential contamination
from the SN host galaxies, and poorly sampled light curves.

The first Hubble diagram from SNe Ia was presented by \citet{kowal68} based on 16 objects with
photographic photometry (before the recognition of the Ia subclass), yielding
a dispersion of $\sigma$=0.61 mag, significantly less than the previous studies,
and an average $M_{pg} = -18.6  + 5 log (H_0/100)$.

In the early 70s, \citet{pskovskii71} detected that, despite their overall similarity, SNe I
have differences in the post-maximum rate of decline. Based on an increased data set of SN I 
lightcurves obtained in the course of the Asiago Supernova Program, 
\citet{barbon73} also noted that it was difficult to fit all SN I photometry with a unique template lightcurve, and
admitted the possibility of the existence of two subclasses: ``fast'' and ``slow''.
\cite{pskovskii77} went a step further, showing a range of a factor of two in the decline
rate of SNe I, and a possible dependence of the Type Ia peak magnitudes with
decline rate, noting that ``the absolute magnitudes of Type I supernovae of junior classes
such as I.6 are 0.9 mag brighter, on the average, than those of class I.14 supernovae''. In 
modern language, \citet{pskovskii77} was reporting that slowly declining SNe were brighter than
rapidly declining objects.

Contrary to the evidence previously reported,
\citet{cadonau85} analyzed a sample of 22 SNe I to construct a mean lightcurve of SNe Ia, 
noting that, contrary to the evidence previously reported, ``individual SNe I show generally no systematic deviations from the templet
light curves'' and that ''the peak luminosity of absorption-free SNe I is also uniform with an intrinsic rms scatter of $<$0.3 mag.''
\citet{leibundgut90} went a step further and established a Hubble diagram based on 35 SNe Ia,
claiming ``a luminosity scatter of 0.25 mag or less''.

The controversy about the uniformity of SNe Ia was finally settled by \citet{phillips93} based
on 9 SNe Ia with modern photometry --mostly from the ``Tololo Supernova Program''-- which indicated 
a range of $\sim$2 mag in absolute peak magnitude and a factor of $\sim$2 in decline rates. A clear correlation 
emerged between decline rate and peak luminosity, with SN 1991T and SN 1991bg being the extreme bright-slow 
and faint-fast SNe, respectively, giving the reason to \citet{pskovskii77} and \citet{barbon73}.

A first long-term effort to search for high-z SNe in galaxy clusters was carried out with the
1.5 m Danish telescope at La Silla Observatory between 1986-1988, resulting in the discovery of SN 1998U at z=0.31
\citep{norgaard89}, thus demonstrating the possibility to extend the study of SNe to cosmologically
interesting distances.

The Cal\'an/Tololo (1989-1993) project provided high-quality CCD photometry for 29 SNe Ia in the Hubble
flow (0.01$<$z$<$0.1) carefully accounting for light contamination from the SN host galaxy through
image subtraction. This dataset confirmed the Phillips' relationship and allowed the measurement
of the expansion rate of the universe accurately for the first time. In fact, the widely
accepted value of the Hubble constant ($H_0$=72) by \citet{freedman01} is essentially based on the Cepheid calibration 
with HST of the distances to seven galaxies that were hosts to Type Ia SNe and a SN Ia Hubble diagram established from
26 Cal\'an/Tololo and 10 SNe from the Center of Astrophysics (CfA) program described below. 
The Cal\'an/Tololo work was one of the key ingredients in the measurement of $H_0$
and the basis for measuring cosmological acceleration by both the HZT and SCP where the 
Cal\'an/Tololo lightcurves represented half of the measurements.

Beginning in 1993, a concerted effort was started by astronomers at the CfA 
to collect modern CCD photometry of SNe Ia, mostly discovered by amateurs. A first
paper reporting 22 high-quality SN light curves was presented by \citet{riess98b}, and later
on augmented by 44 SNe Ia \citep{jha06}.

In the summit of this endeavour lie the two high-z SN searches, HZT and SCP, both of which pushed
technology to the limit in order to discover remote SNe Ia that had exploded even before
the formation of the solar system. CTIO telescopes were fundamental for both teams to make the
discovery of acceleration possible.
For a thorough historical account of this discovery and the race between the two
rival teams the reader is refered to the ``The 4\% Universe'' book by \citet{panek11}.

\section{Summary}
\label{section4}

The discovery of acceleration and dark energy arguably constitutes the most revolutionary discovery
in astrophysics in recent years. CTIO played a key role in this amazing discovery through three
systematic surveys organized by staff astronomers: the ``Tololo Supernova Program`` (1986-2000), 
the Cal\'an/Tololo Project (1989-1993), and the ``High-Z Supernova Search Team'' (1994-1998). CTIO's
state of the art instruments also were fudamental in the independent discovery of acceleration by the
``Supernova Cosmology Project'' (1992-1999). Then, the ESSENCE (2002-2007) project run from CTIO provided valuable constraints on the
properties of dark energy. CTIO remains today as a world-class center at the forefront of SN research through
the ``Dark Energy Survey'' (DES), a five year program (2012-2016) carried with DECam (a mosaic of 62 CCDs) 
mounted on the Blanco telescope with the purpose to probe the origin of the accelerating universe
and help uncover the nature of dark energy.

This story would have ended very differently if not for the visionary efforts of
Prof. Federico Rutllant , the Director of the Chilean National Observatory, who took the initiative
in 1958 to visit Gerard Kuiper, Director of Yerkes Observatory and invite him to witness with his
own eyes the advantageous conditions of the Atacama Desert for astronomical exploration. Joint
efforts between astronomers from the U.S. and the University of Chile began immediately in 
the search of a site for a cooperative observatory in Chile. The site survey, under the direction 
of Dr. J\"{u}rgen Stock led to the establishement of the Cerro Tololo Inter-American Observatory 
in northern Chile in 1962, a very successful story of 50 years of US-Chile collaboration. 

\acknowledgements 
I am very grateful to the organizing committee of this symposium for inviting me to present
this invited talk.
I thank the courtesy of the High-Z Supernova Search Team \citep{riess98a} and the 
Supernova Cosmology Project \citep{perlmutter99} for allowing me to show their figures (reproduced
here as Figures \ref{fig5}, \ref{fig6}, and \ref{fig7}). 
A special thank to Nick Suntzeff for a critical review of this paper, and additional
comments by K. Krisciunas and A. Riess.
I acknowledge support provided by the Millennium Center for Supernova Science
through grant P10-064-F funded by ``Programa Iniciativa Cient\'ifica Milenio del 
Ministerio de Econom\'ia, Fomento y Turismo de Chile''. 

{}

\end{document}